\documentclass[pre,aps,twocolumn,superscriptaddress]{revtex4}
\usepackage{epsfig}
\usepackage{bm}
\newcommand{\B}[1]{{\bm{#1}}}
\newcommand{\C}[1]{{\mathcal{#1}}}
\usepackage[latin1]{inputenc}
\newcommand{\Sub}[1]{_{_{\text {#1}}}} %% Sub- for Upper case
\newcommand{\beq}{\begin{equation}}
\newcommand{\eeq}{\end{equation}}
\newcommand{\bea}{\end{eqnarray}}

\def\Re{${\C R}\mkern-3.1mu e$} %% Reynolds number Re text mode
\def\RE{{\C R}\mkern-3.1mu e} %% Reynolds number Re, math mode

\begin{document}
\title{Identification and Calculation of the Universal Maximum Drag Reduction Asymptote  by Polymers in
Wall Bounded Turbulence}
\author{Roberto Benzi}
\affiliation{Dip. di Fisica and INFN, Universit\`a ``Tor
Vergata", Via della Ricerca Scientifica 1, I-00133 Roma, Italy}
\author{Elisabetta Deangelis}
\affiliation{ Dip. Mecc. Aeron., Universit\`a di Roma ``La Sapienza",
Via Eudossiana 18, 00184, Roma, Italy}
\author{Victor S. L'vov}
\affiliation{Dept. of Chemical Physics, The Weizmann Institute
of Science, Rehovot 76100, Israel}
\author{Itamar Procaccia}
\affiliation{Dept. of Chemical Physics, The Weizmann Institute
of Science, Rehovot 76100, Israel}

\begin{abstract}
Drag reduction by polymers in wall turbulence is bounded from above by a
universal maximal drag reduction (MDR) velocity profile that is a log-law, estimated experimentally by Virk as $V^+(y^+)\approx 11.7 \log y^+ -17$.  Here $V^+(y)$ and $y^+$ are the
mean streamwise velocity and the distance from the wall in  ``wall" units. In
this Letter we propose that this MDR profile is an edge solution of the Navier-Stokes
equations (with an effective viscosity profile) beyond which
no turbulent solutions exist. This insight rationalizes the universality of the MDR and provides a maximum principle which allows  an ab-initio
calculation of the parameters in this law without any viscoelastic
experimental input.
\end{abstract}
\maketitle

The mean  streamwise velocity profile in Newtonian turbulent flows in
channel geometries satisfies the classic von-K\'arm\'an ``log-law of the wall"
which is written in wall units as
\begin{equation}
\label{LLW} V^+(y^+) =\kappa_{_{\rm K}}^{-1}\ln y^+ + B\,,  \quad{\rm
for}~ y^+ \gtrsim 30  \ . \label{Karman}
\end{equation}
Here  $x$, $y$ and $z$ are the streamwise, wall-normal and spanwise
directions respectively \cite{79MY}.  The wall units are defined as
follows: let $p'$ be the fixed pressure gradients $p'\equiv -\partial
p/\partial x$, and  $L$ the mid-height of the channel.  Then the Reynolds
number \Re, the
normalized distance from the wall $y^+$ and the normalized mean
velocity $V^+(y^+)$ (which is in the $x$ direction with a
dependence on $y$ only) are defined by
\begin{equation}
\RE \equiv {L\sqrt{\mathstrut p' L}}/{\nu_0}\ , \  y^+
\equiv {y \RE }/{L} \ , \  V^+ \equiv
{V}/{\sqrt{\mathstrut p'L}} \ , \label{red}
\end{equation}
where $\nu_0$ is the kinematic viscosity. The law (\ref{Karman}) is
universal, independent of the nature of the Newtonian fluid; it is one of
the shortcomings of the theory of wall-bounded turbulence that
the von-K\'arm\a'n constant $\kappa_ K\approx 0.436 $ and the
intercept $B\approx 6.13$ are only known from experiments and simulations
\cite{79MY,97ZS}.

One of the most significant experimental findings \cite{75Vir} concerning turbulent drag
reduction by polymers is that  in channel and pipe geometries the velocity profile
(with polymers added to the Newtonian fluid) is bounded
between von-K\'arm\'an's log-law and another log-law which describes the
maximal possible velocity profile (Maximum Drag Reduction, MDR)
\cite{04LPPT,04DCLPPT,04LPPTa,04BDLPT},
\begin{equation}
V^+(y^+) = {\kappa_{_{\rm V}}}^{-1}\ln\left(e\, \kappa_{_{\rm V}}
y^+\right)\, \quad{\rm
for}~ y^+ \gtrsim 10
 \ . \label{final}
\end{equation}
This law, which had been discovered experimentally by Virk (and hence the
notation $\kappa_{_{\rm V}}$), is also claimed to be universal, independent
of the Newtonian fluid and the nature of the polymer additive, including
flexible and rigid polymers \cite{97VSW}. The numerical value of the coefficient
$\kappa_{_{\rm V}}$ is presently known only from experiments, ${\kappa_{_{\rm
V}}}^{-1}\approx 11.7$, giving a phenomenological MDR law in the form \cite{75Vir}
\begin{equation}
V^+(y^+) = 11.7\ln y^+ -17 \ . \label{finalexp}
\end{equation}
For sufficiently high values of \Re~ and concentration of the polymer, the
velocity profile in a channel is expected to follow the law (\ref{final}).
For finite \Re~, finite concentration and finite extension of the polymers
one expects cross-overs back to a velocity profile parallel to the law
(\ref{Karman}),
but with a larger mean velocity (i.e. with a larger value of the intercept
$B$). The position of
the cross-overs are not universal in the sense that they depend on the nature of the polymers and the flow conditions; the cross-overs are discussed in \cite{04BDLPT,04BLPT}.

While we still cannot predict from first principles the parameters in
von-K\'arm\'an's log-law, the aim of this Letter is to identify the MDR log-law as
an edge turbulent state in wall bounded flows, leading to a derivation
 of the parameters appearing in Eq. (\ref{finalexp}) without any viscoelastic  input. The derivation follows from the theory of drag reduction by
polymers that was developed recently, and therefore a short summary of the
main aspects of the theory  is in order.

Wall bounded turbulence in Newtonian fluids is discussed \cite{79MY, Pope} by considering the
fluid velocity ${\B U}(\B r)$ as a sum of its average (over time) and a
fluctuating part:
\begin{equation}
\B U(\B r,t) = \B V(y) + \B u(\B r,t) \ , \quad \B V(y)
\equiv \langle \B U(\B r,t) \rangle \ . \label{split}
\end{equation}
The objects that enter the Newtonian theory are the mean shear $S(y)$, the
Reynolds stress $W(y)$ and the kinetic energy $K(y)$ :
\begin{equation}
S(y)\equiv d V(y)/d y \ , \!\!\!\quad
 W (y)\equiv - \langle u_x u_y\rangle
\ ,\! \!\!\quad K(y) = \langle |\B u|^2\rangle/2 \ .
\end{equation}
In the presence of dilute polymers added to the Newtonian fluid one needs to
complement  these statistical objects with the
dimensionless ``conformation tensor" $\B R(\B r,t)$ which stems from the
ensemble average of the dyadic product of the end-to-end distance
of the polymer chains, normalized by its equilibrium value $\rho_0^2$
\cite{87BCAH,94BE}.  The way that the conformation tensor appears in the
additional stress tensor which appears in the viscoelastic
equations of motion is model dependent; it is different for flexible and
rigid polymers, and it also depends on the actual model of the
polymers. Nevertheless it was shown theoretically \cite{04BDLPT,04BCLLP} that both for rigid and
flexible polymers one can write down eventually the balance equations for
momentum and energy at distance $y$ away from the wall as
 \begin{eqnarray}
\nu(y) S(y) +W(y) &=& p'L \  ,\quad y\ll L \ , \label{finalmom}\\
 \Big[\nu(y)\frac{a^2}{y^2}+ \frac{b\sqrt{K}}{y}\Big]K(y)&=& W(y)S(y) \  ,
\label{finalbalEvis}\\
\nu(y)& \equiv& \nu_0+   C  \nu_p \langle R_{yy}\rangle \ . \label{Eeffvis}
\end{eqnarray}    
In Eq.(\ref{finalmom})  the right hand side is the rate at which momentum
is generated by the pressure head, $W(y)$ is the momentum flux in physical
space towards the wall, and $\nu(y)S(y)$ stands for the Newtonian viscous
dissipation of momentum in addition to the polymer contribution to the
dissipation of momentum. The effective viscosity $\nu(y)$ is given by Eq.
(\ref{Eeffvis}) where $\nu_p$ is the viscosity due to the polymers in the limit of zero
shear and $C$ is a constant of the order of unity. Similarly, in Eq.
(\ref{finalbalEvis}) the first term on the left hand side  is the combined Newtonian and
polymer contributions to the energy dissipation, the second models the inertial energy fluxes, and the
right hand side is the (exact) energy   production rate. The coefficients $a$ and $b$ 
are dimensionless and of the order of unity.

As in the Newtonian case, the balance equations need to be supplemented by a
relation between $K(y)$ and $W(y)$. Rigorously the Cauchy-Schwartz ineqality
leads to $W(y)\le K(y)$; experimentally one find that
\begin{equation}
W(y)=c\Sub V^2 K(y) \ , \label{WK}
\end{equation}
with $c\Sub V$ apparently $y$-independent outside the viscous boundary
layer. To derive the functional form of the MDR \cite{04LPPT} one asserts that the terms
containing $\nu(y)$ in the balance equations
(\ref{finalmom}) and (\ref {finalbalEvis}) overwhelm the inertial terms, and
then together with (\ref{WK}) one derives $S(y)\sim {\rm Const}./y$ which is the
log-law for the MDR. Consisten with this law the effective viscosiy turns to be linear in $y$. The
increase in the viscosity of course increases the dissipation, but it was argued that the momentum
flux $W$ is reduced even further, leading to an increase in the mean momentum of the flow (for a given pressure head) , i.e.
drag reduction. It is easy to argue \cite{04LPPT} that the slope of the new log-law is
larger that the slope in von-K\'arm\'an's log-law, and hence drag reduction
is obtained. Nevertheless, since neither $c\Sub V$ nor the constants $a$ and
$b$ in (\ref{finalbalEvis}) are known apriori, the actual slope of the MDR
could not be determined. This shortcoming is remedied in the rest of this
Letter.

The crucial new insight that will explain the universality of the MDR and furnish the basis for its calculation is that the MDR  is a marginal flow state of wall-bounded turbulence. In other words this is the solution of
Eqs. (\ref{finalmom}) and (\ref{finalbalEvis}) for which $S(y)$ (or equivalently, the velocity profile) is the maximal possible for any turbulent solution. Attempting to increase $S(y))$ further results in the collapse
of the turbulent solutions in favor of a stable laminar solution $W=0$. As such, the MDR is
universal by definition, and the only question is whether a polymeric (or other additive) can
supply the particular effective viscosity $\nu(y)$ that drives Eqs. (\ref{finalmom}) and (\ref{finalbalEvis}) to attain the marginal solution that maximizes
the velocity profile. We predict that the same marginal state will exist in numerical solutions of the Navier-Stokes equations furnished with a $y$-dependent viscosity $\nu(y)$. There will be no turbulent
solutions with velocity profiles higher than the MDR.

\begin{figure}
\centering \epsfig{width=.50\textwidth,file=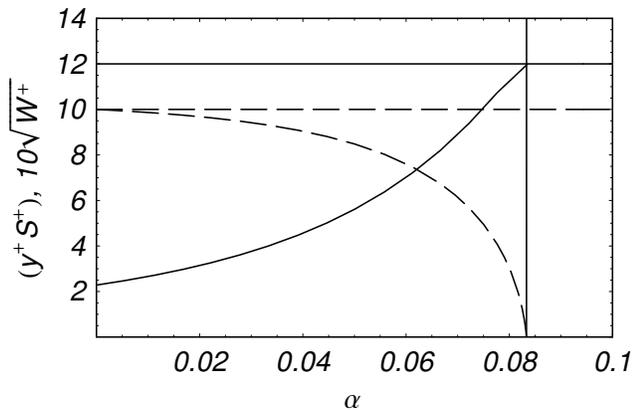}
\caption{The solution for 10$\sqrt{W^+}$ (dashed line) and $y^+S^+$ (solid line) in the asymptotic region $y^+\gg \delta^+$,
as a function of $\alpha$. The vertical solid line $\alpha=1/2\delta^+=1/12$ which is the edge of turbulent
solutions; Since $\sqrt{W^+} $ changes sign here, to the right of this line there are only laminar states.
The horizontal solid line indicates the highest attainable value of the slope of the MDR logarithmic
 law $1/\kappa_{_{\rm V}}=12.$         } \label{Fig1}
\end{figure}

To see this explicitly, we first rewrite the balance equations in wall units. 
Define ${\delta^+}^2\equiv  a^2K(y)/W(y)$, taken for simplicity as $y$-independent; we know
from the Newtonian limit (in which $\nu(y)=\nu_0$) that $\delta^+\approx 6$ \cite{05LPT}. Once we write the equations with $\nu(y)$ the ratio $a^2K/W$ can change drastically, and we
denote it below by ${\Delta}^2$.  With $S^+\equiv
S\nu_0/p'L$, 
$K^+\equiv K/p'L$, $W^+\equiv W/p'L$ and $\nu^+(y^+)\equiv \nu(y^+)/\nu_0$, The balance equations are written as
\begin{eqnarray}
&&\nu^+(y^+) S^+ (y^+) +W^+(y^+)  =1\ , \label{bal1}\\
&&\nu^+(y^+)\frac{{\Delta}^2}{{y^+}^2
}+\frac{\sqrt{W^+}}{\kappa_{_{\rm K}} y^+} =S^+\  . \label{bal2}
\end{eqnarray}
In Eq. (\ref{bal2}) $\Delta\to \delta^+$ when $\nu^+(y^+)\to 1$ (the Newtonian limit). The bunch of numerical constants in the second term on the LHS of (\ref{bal2}) was replaced with  $\kappa_{_{\rm K}}^{-1}$ in agreement with newtonian log-law when $\nu^+(y^+)\to 1$. In fact the second term on the left hand side of Eq.(\ref{bal2}) (which vanishes at the MDR) contains a factor $(\Delta/\delta^+)^3$. This factor is omitted for simplicity;  accounting for this factor complicates slightly the algbra, leaving the final conclusions unchanged. Substituting now
$S^+$ from Eq. (\ref{bal1}) into Eq. (\ref{bal2}) leads to a quadratic equation for $\sqrt{W^+}$. This
equation has as a zero solution for $W^+$ (laminar solution) as long as $\nu^+(y^+){\Delta}/y^+=1$.
Turbulent solutions are possible only when $\nu^+(y^+){\Delta}/y^+ <1$. Thus at the
edge  of existence of turbulent solutions we find $\nu^+\propto y^+$.
This is not surprising, since it was observed already in previous work that the MDR solution is
consistent with an effective viscosity which is asymptotically linear in $y^+$ \cite{04LPPT,04DCLPPT}. It is therefore sufficient
to seek the edge solution by maximizing the velocity profile with respect to linear viscosity
profiles, and we rewrite
Eqs. (\ref{bal1}) and (\ref{bal2}) with an effective viscosity that depends linearly 
on $y^+$ outside the boundary layer of thickness $\delta^+$:
\begin{eqnarray}
&&[1+\alpha(y^+-\delta^+)]S^+ +W^+ =1\ ,\label{pr1}\\
&&[1+\alpha(y^+-\delta^+)]\frac{{\Delta}^2(\alpha)}{{y^+}^2
}+\frac{\sqrt{W^+}}{\kappa_{_{\rm K}} y^+} =S^+\  . \label{pr2}
\end{eqnarray}
We now endow $\Delta$ with an explicit dependence on the slope of the effective
viscosity,  $\Delta(\alpha)\to \delta^+$ when $\alpha\to 0$. For
$\alpha\ne 0$ the ratio $a^2 K/W$ is expected to depend on $\alpha$ (drag reduction involves
a reduction in $W$), and this dependence is an important
part of the following theory. We can present $\Delta(\alpha)$  in terms of
a dimensionless scaling function $f(x)$,
\begin{equation}
\Delta(\alpha) =\delta^+ f(\alpha\delta^+) \ . \label{scaling}
\end{equation}
Obviously, $f(0)=1$. In the Appendix we show that the balance equation (\ref{pr1}) and
(\ref{pr2}) (with the prescribed form of the effective viscosity profile) have an non-trivial
symmetry  that leaves them invariant under rescaling of the wall units.  This symmetry
dictates the function $\Delta(\alpha)$ in the form
\begin{equation}
\Delta(\alpha) =\frac{\delta^+}{1-\alpha\delta^+} \  . \label{Delta}
\end{equation}
Armed with this knowledge we can now find the maximal possible 
velocity far away from the wall, $y^+\gg \delta^+$. There the balance equations
simplify to 
\begin{eqnarray}
&&\alpha y^+S^+ +W^+ =1\ ,\label{pr12}\\
&&\alpha \Delta^2(\alpha)+
\sqrt{W^+}/\kappa_{_{\rm K}}  =y^+S^+\  . \label{pr22}
\end{eqnarray}
These equations have the $y^+$-independent solution for $\sqrt{W^+}$ and
$y^+S^+$:
\begin{eqnarray}
\sqrt{W^+} &=&-\frac{\alpha}{2\kappa_{_{\rm K}}} +\sqrt{\Big(\frac{\alpha}{2\kappa_{_{\rm K}}}\Big)^2
+1-\alpha^2\Delta^2(\alpha)} \ , \nonumber\\
y^+S^+&=&\alpha \Delta^2(\alpha)+\sqrt{W^+}/\kappa_{_{\rm K}} \ . \label{quadsol}
\end{eqnarray}
Obviously, (see Fig. \ref{Fig1}), the supremum of $y^+S^+$ is obtained when $W^+$ vanishes,
which happens precisely when $\alpha=1/ \Delta(\alpha)$. Using Eq. (\ref{Delta}) we find the
solution $\alpha=\alpha_m =1/2\delta^+$.  Then $y^+S^+=\Delta(\alpha_m)$, giving
$\kappa^{-1}_{_{\rm V}}=2\delta^+$. Using the estimate $\delta^+\approx 6$ we get the final
prediction for the MDR. Using Eq.  (\ref{final}) with $\kappa^{-1}_{_{\rm V}}=12$, we get
%%%%%%% FIGURE 1 %%%%%%%%%%%%%%%%%%
%%%%%%%%%%%%%%%%%%%%%%%%%%%%%%%%%%
\begin{equation}
V^+(y^+) \approx 12\ln{y^+}  -17.8 \ . \label{predicta}
\end{equation}
This result is in close agreeement with the empirical law (\ref{finalexp}) proposed by
Virk. Note that the numbers appearing in Virk's law correspond to
$\delta^+= 5.85$, which is well within the error bar on the value of this
Newtonian parameter. Note that we can easily predict where the asymptotic law
turns into the viscous layer upon the approach to the wall. We can consider an 
infinitesimal $W^+$ and solve Eqs. (\ref{bal1}) and (\ref{bal2}) for $S^+$ and the viscosity profile.
The result, as before, is $\nu^+(y)=\Delta(\alpha_m) y^+$. Since the effective viscosity cannot
fall bellow the Newtonian limit $\nu^+=1$ we see that the MDR cannot go below 
$y^+=\Delta(\alpha_m)=2\delta^+$. We thus expect an extension of the viscous layer by a 
factor of 2.

One should note that the result $W^+=0$ should not be interpreted as $W=0$. The differeence
between the two objects is the factor of $\RE^2$, $W\propto \RE^2 W^+$. Since the MDR is reached
asymptotically as $\RE\to \infty$, there is enough turbulence at this state to stretch the polymers to
supply the needed effective viscosity. Nevertheless our discussion is in close correspondence
with the experimental remark by Virk \cite{75Vir} that close to the MDR asymptote the flow appears
laminar.

In summary, one can probably improve further the model of the
Newtonian wall-bounded flow, making it more elaborate and more precise. But the 
message of this Letter will stay unchanged;  whatever is the model of choice, once
endowed with effective viscosity $\nu(y)$ instead of $\nu_0$, there would exist a profile
of $\nu(y)$ that would result in a maximal possible velocity profile at the edge of 
existence of turbulent solutions. That profile is the prediction of the said model of choice for
the MDR. In particular we offer a prediction for simulations: direct numerical simulations of the Navier-Stokes equations in a channel, endowed with a linear viscosity profile \cite{04DCLPPT}, will not be
able to support turbulent solutions when the slope of the viscosity profile exceeds the
critical value that is in correspondence with the slope of the MDR. Notwithstanding, it is gratifying to discover that even a simple model of the balance of energy and momentum is sufficient, in light of the insight presented in this Letter, to predict ab-initio the functional form {\em and} the parameters that determine the Maximum Drag Reduction asymptote.

\appendix
\section{The scaling function}
Consider the following identity:
\begin{eqnarray}
\nu^+(y^+)&=&1+\alpha(y^+-\delta^+)\nonumber\\ &=&[1+\alpha
(y^+-\tilde \delta)+\alpha (\tilde \delta-\delta^+)]\nonumber\\
&=&g(\tilde \delta) \left[1+\frac{\alpha}{g(\tilde \delta)} (y^+ - \tilde \delta)\right] \ ,
\end{eqnarray}
where 
\begin{equation}
g(\tilde \delta)\equiv 1+\alpha(\tilde \delta - \delta^+)\ , \quad \tilde \delta\ge \delta^+ \ . \label{gt}
\end{equation}  
Next introduce newly renormalized units using the effective
viscosity $g(\tilde \delta)$, i.e.
\begin{equation}
y^\ddag\equiv \case{y^+}{g(\tilde \delta)} , \quad \delta^\ddag \equiv \case{\tilde \delta}{g(\tilde \delta)} \ , \quad S^\ddag \equiv S^+ g(\tilde \delta) \ , 
\quad W^\ddag \equiv W^+  \ . \label{vispr}
\end{equation}
In terms of these variables the balance equations are rewritten as
\begin{eqnarray}
&&[1+\alpha(y^\ddag-\delta^\ddag)]S^\ddag+ W^\ddag =1\ , \label{vepr1}\\
&&[1+\alpha(y^\ddag-\delta^\ddag)]\frac{{\Delta}^2(\alpha)}{{y^\ddag}^2}
+\frac{\sqrt{W^\ddag}}{\kappa_{_{\rm K}} y^\ddag} =S^\ddag \ . \label{vepr2}
\end{eqnarray}
These equations are isomorphic to (\ref{pr1}) and (\ref{pr2}) with
$\delta^+$ replaced by $\delta^\ddag$. The ansatz
(\ref{scaling}) is then replaced by
\begin{equation}
\Delta(\alpha) = \frac{\delta^+}{g(\tilde \delta)} f(\alpha \delta^\ddag) \ . \label{Delta2}
\end{equation}
This form is dictated by the following considerations: (i) $\Delta(\alpha)\to \delta^+$ when
$\alpha\to 0$, (ii) all lengths scales in the rescaled units are divided by $g(\tilde \delta)$,
and thus the pre-factor in front of $f$ becomes $\delta^+/g(\tilde \delta)$, and (iii) $\alpha
\delta^+$ in Eq. (\ref{bal2}) is now replaced in Eq. (\ref{vepr2}) by $\alpha\delta^\ddag$, leading
to the new argument of $f$.

Since the function $\Delta(\alpha)$ cannot change due to the change of
variables, the function $\Delta(\alpha)$ given by Eq. (\ref{Delta2}) should be identical to 
that given by Eq. (\ref{scaling}):
\begin{equation}
\delta^+f(\alpha\delta^+)=\frac{\delta^+}{g(\tilde \delta)} f(\alpha \delta^\ddag) \ .
\end{equation}
Using the explicit form of $g(\tilde \delta)$ Eq. (\ref{gt}), and 
choosing (formally first) $\tilde \delta=\tilde\delta ^\ddag=0$ we find that 
 $f(\xi)=1/(1-\xi)$. It is easy to verify that this is indeed the solution of the above equation
 for any value of $\delta^\ddag$, and therefore the unique form of Eq. (\ref{Delta}).

\acknowledgments
We thank T.S. Lo and Anna Pomyalov for very useful discussions, and for careful reading of  early versions of the manuscript. This work has been supported in part by the US-Israel Binational
Science Foundation, by the European Comission under a TMR grant, and by the
Minerva Foundation, Munich, Germany.
%%%%%%%%%%%%%%%%%%%%%%%%%%%%%%%%%%%%%%%%%%%%%%%%%%%%%

\end{document}